\documentclass[useAMS,usenatbib]{mn2e}
\usepackage{graphicx}
\usepackage[update,prepend]{epstopdf}
\title[Linear-drifting subpulse sources]{Linear-drifting subpulse sources in 
radio pulsars}
\author[P. B. Jones]{P. B. Jones\thanks{E-mail:p.jones1@physics.ox.ac.uk}\\
University of Oxford, Department of Physics, Denys Wilkinson Building,\\
Keble Road, Oxford OX1 3RH, U.K.}

\begin{document}

\date{}

\pagerange{\pageref{firstpage}--\pageref{lastpage}} \pubyear{}

\maketitle

\label{firstpage}

\begin{abstract}
Analysis of plasma acceleration in pulsars with positive corotational
charge density has shown that any element of area on the polar cap
is bi-stable: it can be in phases either of pure proton emission or of
mixed ions and protons (the ion phase).  Ion-phase zones are
concentrated near the edge of the polar cap, and are a physical basis
for the coherent
radio emission observed as components within the mean pulse profile.
The state of the polar cap is generally chaotic, but organized
linear motion of ion zones in a peripheral band is possible and is
the likely source of sub-pulse drift.  It is shown that several patterns
of limited movement are possible and can account for the varied
phenomena observed including mirror and bi-directional drifting.
\end{abstract}


\section{Introduction}

The phenomenon of subpulse drift in radio pulsars has been observed
extensively for more than four decades.  The most complete published
survey, that of Weltevrede, Edwards \& Stappers (2006), found it to be
present in about one half of a sample of 187 pulsars, selected only
by signal-to-noise ratio. Usually, subpulses within a window defined
by the averaged pulse profile are observed to drift in successive periods
to either more negative or positive longitudes.  But extremely complex
behaviour has been seen in some individual pulsars.  The direction of
drift can change from negative to positive, or vice-versa. Bi-directional
drifting, in which two or more subpulses within the window of observation
simultaneously drift in opposite directions, has been seen in a small
number of multi-component pulsars, specifically in B1839-04
(Weltevrede et al 2006) and in J0815+09 (Champion et al 2005).  Very
irregular patterns of drift are seen in the multi-component B0826-34
in which the rates change sign smoothly over sequences of the order of
10 periods in length (Gupta et al 2004).  There are many other cases of
complex behaviour such as B0818-41 (Bhattacharyya et al 2007,
Bhattacharyya, Gupta \& Gil 2009).  Also, it is
thought likely that the close association of subpulse drift with
mode-changes and nulls, for example in B1918+19
(Rankin, Wright \& Brown 2013) will be an important diagnostic of
polar-cap physics and radio-emission, particularly in pulsars within
the age range $1-10$ Myr.

Very many authors have sought to describe subpulse motion in terms of the
classic model introduced by Ruderman \& Sutherland (1975) in which localized
regions of electron-positron pair production, referred to as sparks, move
in a circular ${\bf E}\times{\bf B}$ drift around the polar cap.
There has been some limited development of this 
by Deshpande \& Rankin (1999), van Leeuwen \& Timokhin (2012),
also by Gil, Melikidze
\& Geppert (2003) who assumed the existence of a temperature-dependent ion
component in the pair plasma. Although the basis of the model is an
electric-field boundary condition ${\bf E}_{\parallel}\neq 0$
at the polar-cap surface, it is widely used
phenomenologically with no reference to boundary conditions. (The subscripts
$\parallel$ and $\perp$ refer to directions locally parallel with and
perpendicular to the magnetic flux density ${\bf B}$.) Also neglected
is the sign of ${\bf \Omega}\cdot{\bf B}$ at the polar cap,
where ${\bf \Omega}$ is the rotation spin.

In this paper, and based on cohesive energy calculations (see Jones 1985,
Medin \& Lai 2006), it is assumed that the ${\bf E}_{\parallel}\neq 0$
boundary condition is not satisfied except possibly in the extremely small
fraction of pulsars known to have polar-cap dipole fields of the order
of $B\sim 10^{14}$ G.
But spin-down rates measure only the dipole moment and we have to bear
in mind that actual polar magnetic flux densities could be significantly
increased in the presence of higher multipoles.  Thus 
recent numerical modelling of internal field evolution (Geppert,
Gil \& Melikidze 2013) has shown that Hall drift, given suitable initial
conditions, can lead to the formation of small-area magnetic anomalies,
close to the magnetic pole, with the required values of $B$. Geppert
et al, who also comprehensively survey earlier work on Hall drift,
discuss the likelihood that these initial conditions are valid. The
most significant is the existence of
a strong ($\sim 10^{15}$ G) internal toroidal field component within
the crust which must extend to regions close to the poles.  In an 
axisymmetric example of these authors' calculations, Hall drift
amplifies an initial 
$10^{13}$ G dipole field by almost an order of magnitude in $10^{6}$
yr; it also promotes pair creation by decreasing the flux-line radius
of curvature by between
one and two orders of magnitude.  But we return to this question
in Section 5.

Comparisons of the model with observational results have been published
for many pulsars.  A list, which is not complete, includes B0943+10
(Deshpande \& Rankin 2001), B0818-13 (Janssen \& van Leeuwen 2004),
B1857-26 (Mitra \& Rankin 2008), J1819+1305 (Rankin \& Wright 2008),
B1133+16 (Honnapa et al 2012).  In certain cases, the presence of some
evidence for periodicity has lead to the estimation of
a carousel circulation time $\hat{P}_{3} = nP_{3}$ in the usual
notation, where $n$ is here the number of sparks or subpulses and
$P_{3}$ is the observed time interval between the appearance of successive
subpulses at a given longitude.  The model predicts
a unique direction of ${\bf E}\times{\bf B}$ drift so that the
observed cases of bi-directional sub-pulse drift are assumed to arise from
the presence of aliasing, a change from one Nyquist zone to the adjacent.
A positive feature of the model is that the field ${\bf E}$ and hence
the drift velocity can be estimated. However, the complexity and heterogeneity
of subpulse drift and of nulls and mode-changes is perhaps an indication
that physics beyond that of electromagnetic fields and boundary conditions
is involved.

In previous work, we have shown that in the ${\bf \Omega}\cdot{\bf B} < 0$
case, and with the space-charge-limited flow boundary condition, the
blackbody radiation field of the neutron star interacting with accelerated
ions creates a reverse electron flux incident on the polar cap. The
electromagnetic showers formed are a source of protons through the decay
of the giant dipole state and their estimated diffusion time to the surface
is of the order of a typical neutron-star rotation period. It was shown that
zones of ion and proton emission are a physical basis for subpulse
formation and radio-frequency emission through the growth of a two-beam
instability.

An important question is whether or not there is any observational
evidence for the existence of an accelerated ion-proton plasma. Recent
operation of LOFAR by Hassall et al (2012) has placed severe constraints
on the size and altitude of the 40 MHz emission from several pulsars and
from PSR B1133+16 in particular.  On the basis of these results, it has
been argued elsewhere (Jones 2013a) that an electron-positron plasma
cannot be the source of emission in this pulsar.

We refer to Jones (2013b) and earlier papers
cited therein for further details of the model.  
In this paper, Section 2 gives a brief summary
of the properties of the ion and proton zones on the polar cap. 

The reverse electron flux has the same effect as pair creation in limiting
the potential difference available for ion and proton acceleration.  Thus
the outward moving plasma consists of relativistic, but not
ultra-relativistic, ions and protons with negligible pair creation. The loss
of acceleration field ${\bf E}_{\parallel}$
limits the area of the active polar cap in the way described by Arons \&
Scharlemann (1979).  Unfavourably curved
magnetic flux lines from the inactive sector, as defined by
Arons \& Scharlemann, form a dead volume in the open
magnetosphere  of pulsars that are unable to support pair creation.
It is not unreasonable to assume, as in this paper, that the magnetic
flux lines in the region of radio emission are simply those of a dipole.
But the real problem is that the shape of the active polar cap is determined
by the field distribution near the light cylinder which is not well
understood, particularly in the absence of pair creation
(see Muslimov \& Harding 2004, 2009).

The loss of polar-cap symmetry through the presence of the dead volume and
the results of further running of the polar-cap model described in Jones
(2013b) both suggest that the carousel model is not actually realized.
There is, of course, no observational evidence that demands subpulse motion
on a closed curve; only evidence of drift within the band of flux lines
that are observable.
Given our physical model for subpulse formation, it is possible to
envisage much more simple systems of drift that are able to produce
quite naturally the complex forms of behaviour described above. The
computational results from the model are described in Section 3 and
the proposed form of drifting in Section 4.
Our conclusions are given in Section 5.

\section{Basis of the subpulse model}

A summary of the properties of our model can start with proton formation
in the electromagnetic showers produced by the reverse photo-electron flux.
A Green function describes proton diffusion from the point of formation at
the shower maximum to
the upper extremity of the ion atmosphere which we assume to exist in local
thermodynamic equilibrium (LTE) at the polar-cap surface.  The structure of
the atmosphere is defined by the ions which are more numerous than the 
protons.
The protons are never in static equilibrium and a small electric field
present within the charge-neutral atmosphere drives them through to its
upper extremity.  Thus we adopt a $\delta$-function approximation for the
Green function so that a shower formed at time $t$ produces protons at
$t+\tau_{p}$ at the base of the acceleration region.  It has been
estimated that $\tau_{p}$ is of the order of one second which is
a typical pulsar period $P$.
Protons at the top of the atmosphere are preferentially accelerated.
If the rate of proton production exceeds the Goldreich-Julian flux,
$\rho_{GJ}c/{\rm e}$,  the
excess protons form a layer with number density $n_{p}$ at the top of
the LTE ion atmosphere and
acceleration of ions ceases. With our simple Green function, this occurs
at a time $\tau_{p}$ after the start of ion acceleration.

It has to be emphasized that the process described above is quite local.  It
must be true that both accelerated ions and reverse electrons experience
${\bf E}\times{\bf B}$ drift but, because the time constant $\tau_{p}$
is many orders of magnitude longer than particle transit times,
it has negligible effect on the evolution of the system. Any
element of area on the active polar cap is either in a proton zone or
an ion zone according to the character of the accelerated plasma. Our
Green function is a useful but by no means accurate approximation so
that the ion zone will, in general, have both proton and ion components.
This, of course, is necessary if the two-beam instability is to exist
and result in coherent radio emission. We refer to Buschauer \& Benford
(1977) for the relativistic Penrose condition which, in the general case,
governs its growth rate.
But proton zones have only a single component because there is an
accumulation of protons at the top of the LTE atmosphere and there is
no possibility of their contributing to coherent emission.

In the model, the number of protons formed per unit area in an ion zone
of duration $\tau_{p}$ is sufficient to support a proton zone of duration
$\tilde{K}\tau_{p}$ at the coordinates ${\bf u}$  of a
point on the active polar cap,  $\tilde{K}$ being a function primarily
of the surface nuclear charge $Z_{s}$ at ${\bf u}$, the acceleration
potential difference existing on flux lines leaving the surface
at ${\bf u}$, and of the neutron-star
blackbody temperature $T_{s}$ averaged over a fairly large surface area,
extending beyond the open polar cap and subtending possibly as much as
a steradian.

Successive values of $\tilde{K}$ at a given point therefore fluctuate
within a large interval.
The shape and disposition of ion zones is chaotic but not random. From
a given initial state the condition of an element of area at
${\bf u}$ on the polar cap at some future time depends on the sequence
of previous values
of $\tilde{K}$  at all points on the polar cap.  This in turn depends
on the history of the acceleration potential over the entire polar cap.
The time constant $\tau_{p}$ is so long compared with particle transit
times that the acceleration potential can be obtained from Poisson's
equation, with fixed boundary conditions.  We emphasize again that these
considerations, and not ${\bf E}\times{\bf B}$ drift, determine the 
shape-change and movement of an ion zone.

\section{Model results and subpulse drift}

A complete description of the way in which computational results have
been obtained from the model polar cap is contained in Jones (2013b).
However, we shall summarize here those features that are of immediate
interest.

The polar cap is assumed circular and initially divided into
$n_{s} = N^{2}$ elements of equal area
arranged in $N$ concentric annuli, the central element being a
circle of radius $u_{0}/N$, where $u_{0}$ is the polar-cap radius,
\begin{eqnarray}
u_{0} = \left(\frac{2\pi R^{3}}{cPf(1)}\right)^{1/2}.
\end{eqnarray}
In this, $P$ is the rotation period, $R = 1.2\times 10^{6}$ cm is the
neutron-star radius and $f(1) = 1.368$ for a $1.4$ $M_{\odot}$ star
(Muslimov \& Harding 2001).
The polar-cap magnetic flux density is $3\times 10^{12}$ G and the
selected periods $P = 1,2,3$ s.

In the initial state, all elements are proton zones. Selected in a
random sequence, and with randomly chosen time intervals, all elements
make the transition to the ion phase.  After each transition,
self-consistent solutions are obtained for all ion zones (the complete
polar cap) giving
individually the degree of ionization of the ions, the energy flux of
reverse photo-electrons, and the acceleration potential on the cell
axis.  Each transition back to the proton phase occurs after an interval
$\tau_{p}$ at which time the duration $\tilde{K}\tau_{p}$ of the
subsequent proton phase has been calculated.  After each transition
from an ion to a proton phase there is
a further solution for self-consistency of the complete polar cap.
The model is elementary in character but has the considerable advantage
that each set of calculations made between the transition to the ion
phase in a particular element and its reverse at a time $\tau_{p}$
later is self-contained.  Thus the program can be allowed to run for
as long a time as is convenient without the accumulation of errors.

The more significant results of such runs, for $N = 10$ were the extent
of polar-cap potential fluctuations on time-scales of the order of 
$\tau_{p}$ and that, at any instant, most ion-phase elements are
positioned near the boundary at ${\bf u}\approx {\bf u}_{0}$.  The
latter is entirely
consistent with the two-component emission structure (described as
conical) observed in many pulsars, and is an anticipated consequence
of the variation of the time-averaged value of $\tilde{K}$ as a
function of ${\bf u}$. Computation of the autocorrelation function
for the azimuthal distribution of
ion-zone elements near ${\bf u}_{0}$ also gave evidence of the
preferential formation of small clusters which could be the
source of sub-pulses. Although
obtained for a circular polar cap, these results are qualitatively
independent of polar-cap shape and would remain valid for the
approximate form we assume in the following Section.
 
A modified version of the program with $N = 10$ has the $2N-1 =19$
elements of the outer annulus increased to 60 in order to facilitate
the search for naturally occurring subpulse drift similar to the
carousel model of Ruderman \& Sutherland.  In a further optional
modification, all elements except those in the outer annulus are
at all times in the proton phase so that the symmetry of the
acceleration potential about the magnetic axis is disturbed only by
ion-phase elements in the outer annulus.  Even with this unrealistic
structure, attempts to see if subpulses formed in the outer annulus
with an initial drift settle down to stable and continuous long-term
drifting have met with no success.  Subpulses prove to be unstable after
no more than several complete rotations.  In itself, this result is
unremarkable and may be no more than a consequence of the limitations
of the program.  But the significant result is that if elements in
the inner annuli are permitted to undergo transitions to and from
the ion phase, any stable subpulse structure there is in the outer
annulus is destroyed almost immediately.

\begin{figure}
\includegraphics[trim=10mm 60mm 10mm 90mm, clip,width=84mm]{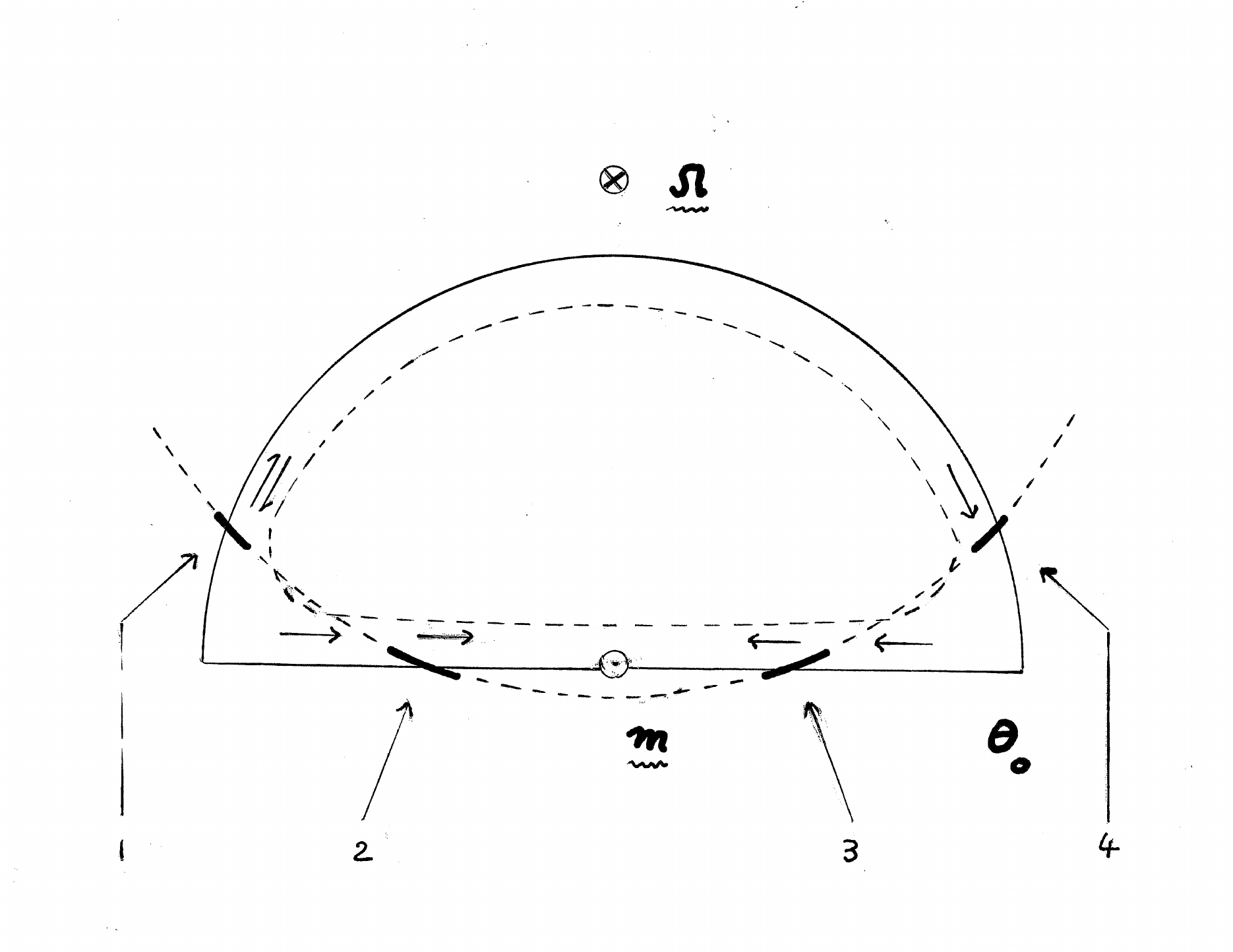}
\caption{The polar diagram with the magnetic axis ${\bf m}$ as origin 
shows the line of sight, spin ${\bf \Omega}$, and the tangents to
the active open magnetosphere at altitude $\eta = 10$ and polar angle
$\theta_{0} = 4.6^{\circ}$.
The peripheral band of flux lines supporting emission
is shown schematically as a closed broken curve.  Relative to the
magnetosphere, the line of
sight transits on a curve shown by the broken line whose intersections
with the peripheral band give four components within the mean pulse
profile.  Arrows show the directions ion-zone drift that correspond
with the observed pattern of B0815+0939.  The transit curve is drawn
approximately to scale so as to reproduce the $100^{\circ}$ mean
profile width of this pulsar, and is at an angle of $\sim 6^{\circ}$
with $-{\bf \Omega}$.  The components seen are labelled $1-4$ in order
of increasing longitude, the first being described as indistinct
by Champion et al.}
\end{figure}

The presence of chaotic ion-zone regions in the central part of the
polar cap is to be expected and cannot be simply an artefact arising
from the limitations of our model and program. Hence it is difficult
to see how the degree of stability necessary for long-term carousel
motion can exist.

In Section 1, we noted that some authors have been able to find what
is interpreted as a carousel rotation time by measurements of
$P_{3}$ combined with estimates of the integer number of sparks,
or by two-dimensional
longitude-resolved fluctuation power spectra formed from long
sequences of successive pulses (see, for example, Rankin et al 2013).
But failing direct observation of the whole polar cap, these can be
no more than evidence for a periodicity.  The following Section
introduces a simple linear drift pattern which is less demanding of
long-term stability and appears more able to accommodate the sui
generis nature of many pulsars.

\section{Patterns of linear drift}

Before proceeding to a description of the linear drift patterns, it
is necessary to  mention briefly those factors that make the
generation  of coherent radio emission possible.
The exponential amplitude growth factor $\exp\Lambda$ for the two-beam
instability assumed to be the source is given by,
\begin{eqnarray}
\Lambda \approx 2.4\times 10^{5}\left(\frac{B_{12}mZ_{\infty}}
{PM\gamma^{3}_{A,Z}}\right)^{1/2}\left(1 - \eta^{-1/2}\right)
\end{eqnarray}
at radius $r = \eta R$.  Here $B_{12}$ is the polar-cap magnetic flux
density in units of $10^{12}$ G, $m$ is the electron rest-mass, and
$Z_{\infty}$ is the final charge of the accelerated ion. The ion mass
is $M$, $\gamma_{A,Z}$ is its Lorentz factor, and $P$ is the rotation
period in seconds.  The exponent $\Lambda$ is a slowly
varying function of the proton fraction of the polar-cap current
density which we here assume to be $0.2$.  We
refer to Jones (2012) for the derivation of this expression.

Equation (2) shows that much of the gain in amplitude occurs at
altitudes, $\eta < 2$, beyond which some further
acceleration through the Lense-Thirring effect
(Muslimov \& Tsygan 1992) can occur in principle unless blocked by
continuing photo-ionization. But the larger part of the gain is at
altitudes above those where most
of the acceleration occurs. These considerations are significant
because, as equation (2) demonstrates, the exponent $\Lambda$ is
dependent on the ion Lorentz factor. Amplitude gains can be very large
for ions that are accelerated from regions ${\bf u}$ of the polar-cap
surface close to the the boundary ${\bf u}_{0}$ because they experience
a small ${\bf E}_{\parallel}$ field.  Ions of greater Lorentz factor
have smaller $\Lambda$ so that, given the exponential form of the
gain, a sharp cut-off would be anticipated at some critical
value of $\gamma_{A,Z}$. In general, the model described in Section 3
has a time-averaged acceleration potential difference
which as a function of ${\bf u}$ has broadly the same form as it
would have in the absence of the ion phase, but of reduced magnitude.
Limitation of the polar-cap area by the introduction of a dead zone,
as in the following paragraph, also reduces the
acceleration potential difference by a factor $\zeta$ which is
shape-dependent but satisfies $1/4 < \zeta < 1/2$ for the
semi-circular form assumed.
Even so, we predict that ions from the central regions of the 
polar cap will often be too energetic to give the gain necessary
for the non-linearity and plasma turbulence which are believed to be
the source of the emission.  Thus it is anticipated that, in many
cases, the source of observable emission will be confined to 
a band on the polar-cap surface close to the boundary ${\bf u}_{0}$.

This is shown schematically in Fig. 1 which is a polar diagram
showing the angular relations between the line of sight, spin
${\bf \Omega}$, and the tangents to the active open magnetosphere
flux lines at altitude $\eta$.  If ${\bf u}_{0}$ is here the
boundary of the active polar-cap surface, the tangent at $\eta$ is
at an angle $\theta_{0} = 3\eta^{1/2}{\bf u}_{0}/2R$
to the magnetic axis for
a dipole field. The papers by Muslimov \& Harding (2004, 2009)
indicate that for general values of $\psi$, the inclination of
the magnetic axis to ${\bf \Omega}$, the 
original Arons \& Scharlemann (1979) division of the open
magnetosphere into favourably and unfavourably-curved magnetic
flux lines has some physical basis.  For this reason we assume
a semi-circular polar cap having a radius ${\bf u}_{0}$ given
by equation (1). Also shown is the curve made by the
transit of the open magnetosphere across the line of sight, for a
particular case in which the latter is at a small angle with
${\bf \Omega}$.  This is the $P = 0.645$ s PSR J0815+0939
(Champion et al 2005). The
relation between the transit curve and open magnetosphere is
drawn (approximately) to scale at an altitude $\eta = 10$,
consistent with the work of Hassall et al (2012), so that the
integrated profile has the observed width of $100^{\circ}$. 
The value of $\psi$ is then $\sim 174^{\circ}$, and the number of
components observed within the integrated pulse profile is equal
to the number of times the line of sight crosses the emission
band.  It is seen immediately, in principle, how the four
components of this pulsar arise and it is also possible to
envisage other polar-cap shapes that would give four
intersections per transit.  

However, the specific case chosen here
should not be considered too seriously because it appears to
us that the real shape of the polar cap in cases where $\psi$ is
close to either $0$ or $180^{\circ}$ is particularly obscure.
There appear to be no published papers that have considered
such small inclinations.
But for pulsars with more general values of $\psi$, the origin of
two-component (referred to as single-cone) emission is obvious
for almost any shape of polar cap.
As an example, we refer to B1133+16 which has been observed over
a wide range of frequencies by Hassall et al (2012). Reference to
Fig. 1 shows that the outer and the central components can be
easily understood.  Thus the angular separation of the two outer
components at $181$ MHz, assuming emission at $\eta = 10$, should
be $2\theta_{0}\csc\psi = 8.6^{\circ}$ for the inclination
$\psi = 180^{\circ}- 51^{\circ} = 129^{\circ}$ assumed by
Hassall et al, to be compared with
the observed value of $8.2^{\circ}$.  The observed separation at
the lowest frequency, $48$ MHz, is rather larger ($12.3^{\circ}$)
but we do not regard this discrepancy as serious because this pulsar
does exhibit some degree of radius-to-frequency mapping.
Apart from these geometrical considerations,
it is also easy to see how simple ion-phase motions give
rise to the observed sub-pulse drift patterns, although
a number of complicating factors exist. 

Emission observed on the line of sight does not, in general, originate
from a line on the polar-cap surface because the radiation
decouples from the plasma over a finite interval of altitude.
But there is an additional reason
which is that, quite naturally for the class of mechanism we assume,
the radiation
emitted is not precisely parallel with the local magnetic flux
density.  Referring again to B1133+16, we note that the observed
angular widths of the outer components are not inconsistent with
those values of $k_{\perp}/k_{\parallel}$, where ${\bf k}$ is
the radiation wave-vector, which appear possible for the
quasi-longitudinal Langmuir mode introduced by Asseo, Pelletier \& Sol
(1990) and described in the present context by Jones (2012).
Thus the ions producing radiation on the line of sight
are actually emitted from the intersection of a finite
band on the polar cap with the peripheral band shown in Fig. 1.

A further complication is that, for simplicity, the study of
quasi-longitudinal Langmuir modes by Asseo et al (1990) from
which equation (2) follows is essentially one-dimensional. But
because we are concerned with emission from a peripheral band,
as in Fig. 1, it is necessary to consider the lateral dimension
of the emission region in relation to the observer-frame wavelength
of the radiation.  In general, there is a difference of one or
two orders of magnitude.  However, this may limit the extent to
which ions from close to ${\bf u}_{0}$ with Lorentz factors
near unity can be efficient emitters. But these considerations
do not seriously affect the arguments which follow.

In relation to sub-pulse formation, the program described in
Section 3 (see Jones 2013b) indicates the formation of clusters
of ion-phase elements near the boundary ${\bf u}_{0}$, but with
no long-term stability.  A system of organized ion-zone motion
leading to sub-pulse formation and drift is shown by the
arrows in the peripheral polar-cap band in Fig.1. We
emphasize that organized motion over the whole polar cap is not
needed and may not occur.  The features of ion-zone motion that
lead to this conclusion will be discussed in the following
paragraph. Also, there is usually too little amplitude gain
given by equation (2) for motion in the central region of the
polar cap to be observed.
We emphasize that long-term ion-zone
stability is not needed as it would be in continuous carousel
motion because an individual zone survives for
only a single transit.  Also the number of zones present at a
given instant need not be an integer because one may be in a
state of partial termination at ${\bf u}_{0}$.

\begin{figure}
\includegraphics[trim=10mm 15mm 20mm 160mm, clip, width=84mm]{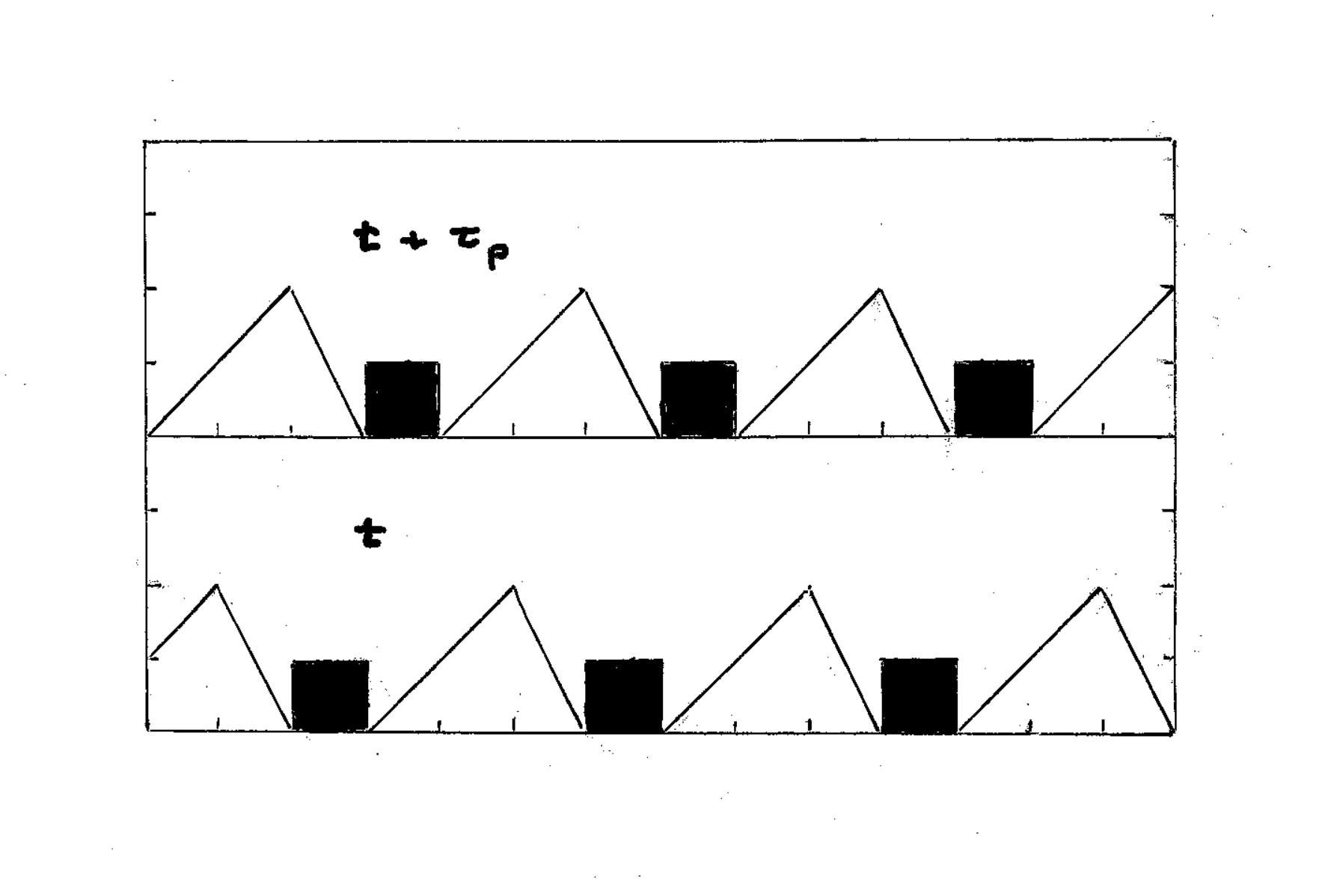}
\caption{Ion-zone motion with velocity $v$ is shown in one
dimension for idealized proton diffusion, the horizontal
axis being distance in units of $v\tau_{p}$.  There are two
vertical scales: the ion-zone current density (solid black)
in units of $\rho_{GJ}c$, and for the line diagram giving
the proton number density $n_{p}$ in units of
$\rho_{GJ}/{\rm e}$. The maximum value of $n_{p}$ is
$\tilde{K} - 1$.  The duration of the proton phase at any
point is $\tilde{K}\tau_{p}$, with $\tilde{K} = 3$ in the
diagram.  An ion zone advances as the proton number density
in front of it is depleted at a rate given by the
Goldreich-Julian flux.
The lower panel is at time $t$ and the upper at
$t + \tau_{p}$.  The zeros in $n_{p}$ remain for the true
proton diffusion Green function, although the detailed
shape of the $n_{p}$-distribution would change.}
\end{figure}

The way in which zones move was described in Section 5.1 of
Jones (2013b), but is shown here in one dimension in Fig. 2,
which assumes the
idealized proton formation Green function mentioned in Section 2
and a value of the parameter $\tilde{K} = 3$.
For either the ideal or the true Green function, motion or
change of shape of any ion-zone boundary in two
dimensions is determined solely by the following factors.

(i)   Any element of area within an ion zone reverts to the
proton phase when proton creation is sufficient to give a
proton number density $n_{p} > 0$ at the top of its LTE
ion atmosphere. For the model described in Sections 2 and 3,
this occurs after a time $\tau_{p}$.

(ii)  Any element of area in a proton zone reverts to the ion
phase when its proton number density falls to $n_{p} = 0$.
 
(iii) The speed and direction of motion of any element of
ion-zone boundary is determined by the local value of
$\nabla_{{\bf u}}n_{p}$. The rate of decrease of $n_{p}$
at any point is constant, determined by the Goldreich-Julian
current density. The direction of motion is then that of
the least gradient.

Time-averaged values of $\tilde{K}$ and hence of $n_{p}$ are
greatest in the central region of the polar cap and are
lower near ${\bf u}_{0}$.  Clearly, movement within the
peripheral band is strongly favoured.  It is essentially
one-dimensional and can be in either direction 
with a velocity $v$ on the polar-cap surface given by
$w = v\tau_{p}$, where $w$ is the zone width.  Proton zones
then are of length $\tilde{K}w$.  In a real system, the zeros
in the surface proton number density at the leading and trailing
edges of an
ion zone would remain, but for the correct Green function, 
both protons and ions are  accelerated during the ion
phase.  We suggest that the velocity is defined by $\tau_{p}$
and by $w$, and that the magnitude of $w$ is determined by
Poisson's equation and by the photo-ionization and
reverse-electron flow processes, that is, the tendency to
cluster as found in the model. A further factor that might
be relevant to the size of $w$ is that electromagnetic
showers also produce neutrons. The probability that a
neutron would survive to reach the top of the LTE atmosphere
is hard to estimate but $(n,\gamma)$ reactions could mediate
processes that are non-local, unlike those described in
Section 2. However, organized motion need
not be universal.  Thus the possible states of the polar-cap
are consistent with classes of coherent or diffuse
drifting seen by Weltevrede et al.

Our model predicts that $P_{3} = (\tilde{K} + 1)\tau_{p}$,
and estimates of $\tau_{p}$, and of $\tilde{K}$ in the peripheral
polar-cap band, give values of the order of $2-3$ s (see
Jones 2013b).  Neither of the two parameters is other than
a slowly-varying  function of the basic pulsar parameters,
$P$, $B$, $P/2\dot{P}$, and this is consistent with the
conclusion of the Weltevrede et al survey.  Measured values
$P_{3}$ are largely within the $1-10$ s interval, which is
relatively compact for pulsar physics, but many authors have
suggested that aliasing is present.  Thus the distribution
of true values
of $P_{3}$ (or of the speed $v$ on the polar cap) may be
even more compact.  Aliasing can be present in the results of
periodic sampling of any system, including the model of this
paper, and is a 
possible explanation for those few relatively large $P_{3}$
that are seen.  But it is not required for bi-directional
drifting.

The organized drift shown in Fig. 1 corresponds with the
mirrored drift patterns observed in J0815+0939 and does so
without the need for aliasing. (We note, however, that the
behaviour of the first drift band seen by Champion et al
is indistinct. This could be a consequence of the direction
of drift being almost perpendicular to the transit curve,
which may also explain the longitude-stationary class of
drift noted by Weltevrede et al.)
Linear motions may result in the
collision of two ion zones moving in opposite directions.
The result of such a collision can be seen by imagining that
the mirror image of Fig. 2 is placed adjacent to its right-hand
side.  The two zones touch after the proton zone separating them
becomes exhausted and at a time $\tau_{p}$ later both terminate,
having effectively annihilated leaving a proton zone. An
ion zone simply terminates on impact with the polar-cap
boundary.

The central region of the polar cap is the source of what is
referred to as core emission.  It may or may not be observable
depending on the rate of growth given by equation (2). In
many instances, the local value of $\gamma_{A,Z}$ will be too
large for the necessary growth.  The region has, in general,
larger values of $\tilde{K}$ than in the peripheral band
and behaviour that is less organized and more chaotic is
anticipated.

\section{Conclusions}

The motivation for this work has been to construct a physical
framework forming a basis for understanding how and why the
many different observed phenomena occur.  The mode changes,
nulls, and sub-pulse drifts are to a large extent individual
in character. It is true that there are
distributions of rotation period, polar magnetic field, and
of the inclinations of the line of sight and magnetic axis to
${\bf \Omega}$.  But we regard it as questionable that
Maxwell's equations with the appropriate boundary conditions
and with the basic processes of quantum electrodynamics
can provide an understanding of the complex phenomena that are
seen in many specific pulsars.

A positive feature of the present model is that
it does not require special values of the
surface magnetic field.  Whilst Geppert et al (2013) have
demonstrated a
field amplification of almost an order of magnitude for an
initial dipole component of $10^{13}$ G, it is not clear
from their paper whether amplification factors of
$10^{2} - 10^{3}$ would be possible to produce fields of
about $\sim 10^{14}$ G in a large fraction of the pulsar
population.
Also, we do not believe that the photo-production
of protons by the reverse electron flux in the ${\bf E}\neq 0$
polar cap has no observational consequences and so
can be neglected.

The proton cohesive energy is negligible, even at $10^{14}$ G,
and therefore an atmosphere would form if the production rate
exceeded the Goldreich-Julian flux.  Accelerated protons have
a negligible pair-production rate, and there is no obvious
mechanism for coherent curvature radiation.  If high-$B$
magnetic anomalies exist in some pulsars and are strong enough
to produce the ${\bf E}_{\parallel}\neq 0$ condition at the
polar cap, 
the proton-emission phase would interrupt this for substantial
intervals of time so that their radio emission would be
intermittent, possibly as in the Rotating Radio Transients (RRAT).

There is no obvious way in 
which the neutron-star surface can be involved in the case
of ${\bf \Omega}\cdot{\bf B} > 0$ pulsars because electrons
are the only negatively-charged particles which can form a
polar-cap current density. But the polar-cap alignment
${\bf \Omega}\cdot{\bf B} < 0$ introduces two
further variables both of which, unfortunately, are not well
known.  These are the nuclear charge $Z_{s}$ of surface ions
and the polar-cap surface temperature $T_{s}$ that is effective
in photo-ionization of accelerated ions. It also introduces
bi-stability in the composition of the accelerated ion beam.

Previous work (Jones 2013b) has demonstrated how this
bi-stability distributed over the polar cap leads to potential
fluctuations, on time-scales of the order of $\tau_{p}$, and
to the formation of localized areas from which the accelerated
plasma has the properties necessary for rapid growth of a
quasi-longitudinal Langmuir mode.  These are the sub-pulses
observed in pulsars with ages $\sim 1-10$ Myr, or more, and we
propose that our model provides a sound physical basis for
their formation. They
do not have the long-term stability needed for a continuous
circulation model, but we have pointed out in Section 4 that such
motion is not needed and that a subpulse need survive only for
the transit of no more than a polar-cap radius.

The physical processes which form the basis of the model are
both prosaic and well-established.  Provided neutron stars with
${\bf \Omega}\cdot{\bf B} <0$ exist, these processes must be
present.  They do not rely on high polar magnetic-flux
densities except for the generation the modest acceleration
potential differences which are needed. In general, there is
no requirement that the basic parameters should have values
within specified intervals and most
neutron stars formed with this alignment should pass through
an epoch of nulls, sub-pulse drift, and possibly mode-changes.
The exception and the particular difficulty of the model is
its dependence on the surface temperature $T_{s}$. This is
the surface temperature of the area which is the
source of the blackbody radiation effective in photo-ionization.
It subtends, perhaps, a steradian  centred on the polar cap and
owing to anisotropy of the thermal conductivity,
is likely to be at a higher temperature than the mean for
the whole star. The model indicates that
the important temperatures
are those within a factor of approximately two of
$4\times 10^{5}$ K 
at the neutron-star surface, or $3\times 10^{5}$ K as seen
by a distant observer.  These are very low, not observable for
a compact object, but can be maintained by quite modest forms of
dissipation. They must be attained during some interval in the
life of the star, but it does not appear possible to estimate
with confidence the length of that epoch.

The model is a physics-based framework but unfortunately the
extent to which it is either capable of prediction or
of being falsified is limited.  In particular, sub-pulse
movement is not easily predictable.
We have attempted to show that organized
motion can occur in a band at the polar-cap perimeter,
and is responsible for the sub-pulse drift observed.
But in general, the state of the entire polar
cap is chaotic.  However, given the heterogeneity of
mode-changes, nulls, and sub-pulse drift, the search
for a simple physical theory may well prove futile.
The most direct evidence for ionic rather than
electron-positron plasma appears to be from emission-altitude
measurements in pulsars with favourable signal-to-noise ratios,
such as those of Hassall et al (2012). Further measurements
of this kind, should they prove possible, will be of
considerable interest.

\section*{Acknowledgments}

The author thanks the referee for extensive comments that have
greatly improved the presentation of this work.

\bsp

\label{lastpage}

\end{document}